\documentclass[aps,prx,twocolumn,groupedaddress,superscriptaddress,showpacs,amssymb,amsmath,amsfonts,floatfix,nofootinbib]{revtex4-2}

\usepackage{mathtools}
\usepackage{perpage} 
\MakePerPage{footnote}
\usepackage{color}
\usepackage{braket}
\usepackage{subfigure}

\usepackage{graphicx}
\usepackage{float}
\usepackage{graphicx}
\usepackage{amsmath}
\usepackage{hyperref}
\usepackage{amsmath}  % usually already loaded by revtex4-2
\usepackage{amssymb}
\usepackage{booktabs}

\begin{document}
	
\title{Parallel Data Processing in Quantum Machine Learning}

\author{Sina Asadiyan Zargar}
\affiliation{Centre for Quantum Engineering and Photonics Technology, Sharif University of Technology, Tehran, Iran}
\affiliation{Department of Electrical Engineering, Sharif University of Technology, Tehran, Iran}

\author{Mehdi Ramezani}
\email[]{mehdiramezani200@gmail.com}
\affiliation{Centre for Quantum Engineering and Photonics Technology, Sharif University of Technology, Tehran, Iran}

\author{Abolfazl Bahrampour}
\affiliation{Centre for Quantum Engineering and Photonics Technology, Sharif University of Technology, Tehran, Iran}

\author{Saeed Bagheri Shouraki}
\affiliation{Department of Electrical Engineering, Sharif University of Technology, Tehran, Iran}

\author{Alireza Bahrampour}
\affiliation{Centre for Quantum Engineering and Photonics Technology, Sharif University of Technology, Tehran, Iran}
\affiliation{Department of Physics, Sharif University of Technology, Tehran, Iran}
	
\date{\today}

\begin{abstract}
	\begin{center}
		\textbf{Abstract}
	\end{center}
	We propose a Quantum Machine Learning (QML) framework that applies the core design principle of quantum algorithms—superposition, oracle, and interference—to accelerate training. Building on the structural analogy between feature extraction in foundational quantum algorithms and parameter optimization in QML, we reformulate the training process to leverage quantum parallelism: all training samples are encoded into a quantum superposition, processed through a parameterized quantum circuit, and classified via an interferometer module that implements quantum interference across the dataset. This architectural reformulation reduces the theoretical complexity of loss function evaluation from $O(N^{2})$ in conventional QML training to $O(N)$, where $N$ is the dataset size. Numerical simulations on multiple binary and multi-class classification datasets (with up to $N=128$ samples) demonstrate that our method achieves classification accuracies comparable to conventional circuits while reducing the number of quantum circuit executions per cost function evaluation from $N$ to 1. This represents a near $N$-fold reduction in quantum overhead per training iteration, reducing the required circuit executions without loss of accuracy. These results highlight the potential of quantum algorithmic design principles as a scalable pathway to efficient QML implementations.
\end{abstract}

\maketitle

%%%%%%%%%%%%%%%%%%%%%%%%%%%%%%%%%%%%%%%%%%%%%%%%%%%%%%%%
%%%%%%%%%%%%%%%%%%%%%%%%%%%%%%%%%%%%%%%%%%%%%%%%%%%%%%%%

\section{Introduction}

Quantum Machine Learning (QML) has emerged as a transformative paradigm, promising to leverage the unique properties of quantum mechanics to solve complex computational problems that are intractable for classical systems. By mapping data into high-dimensional Hilbert spaces, Variational Quantum Classifiers (VQCs) aim to find patterns and decision boundaries with a theoretical efficiency that exceeds classical kernels. As the field moves toward large-scale applications, the focus has shifted from demonstrating “quantum supremacy” to achieving practical scalability on Noisy Intermediate-Scale Quantum (NISQ) devices.

However, existing QML solutions, while promising, suffer from a fundamental architectural inefficiency: they typically process training samples sequentially. In a conventional VQC, evaluating the cost function for a dataset of size $N$ requires $N$ separate quantum circuit executions per training step. When the number of trainable parameters also scales with the complexity of the data, the total training complexity can reach $O(N^{2})$. This quadratic scaling creates a significant “wall-clock” time bottleneck, negating much of the theoretical advantage of quantum acceleration and making the training of large-scale models prohibitively expensive in terms of both time and quantum resource allocation.

\begin{figure}[htb]
	\centering
	\includegraphics[width=0.45\textwidth]{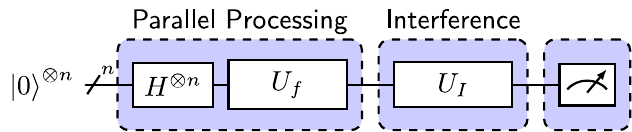}
	\caption{Schematic representation of the shared three-stage architecture underlying foundational quantum algorithms—comprising superposition preparation, oracle-based encoding, and quantum interference}
	\label{fig-quantum-algorithms}
\end{figure}

Building on the structural analogy between feature extraction in foundational quantum algorithms—such as the Deutsch-Jozsa \cite{deutsch1992rapid} and Bernstein-Vazirani \cite{bernstein1997quantum} algorithms—and parameter optimization in QML, we propose a parallelized framework that reformulates the training process. Instead of treating training samples as isolated inputs, our architecture applies the core design principles of quantum algorithms—superposition, oracle-based encoding, and interference (FIG. \ref{fig-quantum-algorithms})—to process the entire dataset simultaneously. By encoding all training samples into a single quantum superposition and applying a parameterized variational ansatz to this global state, we reduce the theoretical complexity of cost function evaluation from $O(N^{2})$ in conventional sequential training to $O(N)$.

The core innovation of this work lies in treating the training dataset as a structured quantum oracle, a concept widely used in quantum query complexity but underutilized in variational QML. The primary reason such an approach has remained elusive is the perceived overhead of state preparation; however, we demonstrate that by integrating index-based encoding directly into the variational loop, the $O(N)$ preparation cost is effectively amortized, maintaining a clear advantage over sequential methods.

To validate this approach, we conducted extensive numerical simulations on multiple binary and multi-class classification datasets, with datasets up to $N=128$ samples. Our results demonstrate that the parallel model achieves classification accuracies comparable to (and in some cases, within 1\% of) conventional sequential circuits.

The remainder of this paper is organized as follows: Section \ref{sec:qml} provides the necessary background on quantum machine learning. Section \ref{sec:method} details the proposed parallel architecture and complexity analysis. Section \ref{sec:SimulationResults} presents the experimental setup and numerical results, and Section \ref{sec:conclusion} concludes with a discussion of limitations and future work.

%%%%%%%%%%%%%%%%%%%%%%%%%%%%%%%%%%%%%%%%%%%%%%%%%%%%%%%%
%%%%%%%%%%%%%%%%%%%%%%%%%%%%%%%%%%%%%%%%%%%%%%%%%%%%%%%%

\section{General Framework of Quantum Machine Learning}\label{sec:qml}

% Introducing the supervised learning framework for quantum machine learning
In quantum machine learning (QML), the goal in supervised learning is to find a hypothesis function $h: \mathcal{X} \to \mathcal{Y}$ that approximates the true labeling function $f^*$ governing input-output pairs $(\mathbf{x}, y) \in \mathcal{X} \times \mathcal{Y}$, sampled from an unknown joint distribution $\mathcal{D}$, i.e., $(\mathbf{x}, y) \sim \mathcal{D}$. The objective is to minimize the expected risk:
\begin{equation}
	\mathcal{L}(h) = \mathbb{E}_{(\mathbf{x}, y) \sim \mathcal{D}}[\ell(h(\mathbf{x}), y)],
\end{equation}
where $\ell$ is a suitable loss function, such as mean squared error or cross-entropy. Parameterized Quantum Circuits (PQCs) have emerged as a flexible and powerful framework for constructing such hypothesis functions in QML, particularly for classification, regression, and generative modeling tasks \citep{Schuld2020Circuit, Kandala2017Hardware, Peters2021Machine}.

\begin{figure}[htb]
	\centering
	\includegraphics[width=0.45\textwidth]{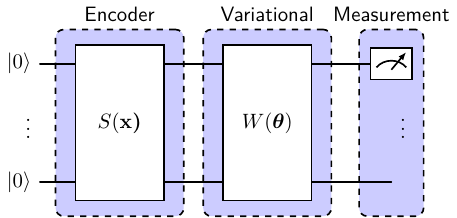}
	\caption{Architecture of a Parameterized Quantum Circuit (PQC) classifier comprising two main components: an input embedding circuit that encodes classical data into quantum states, and a variational ansatz composed of trainable quantum gates. The parameters of the variational layer are optimized using a classical optimizer to minimize a cost function based on the measurement outcomes, enabling the model to learn discriminative patterns in the input data.}
	\label{fig-pqc}
\end{figure}

% Describing the structure and operation of PQCs
PQCs consist of quantum gates with tunable parameters, optimized within a hybrid quantum-classical loop using classical algorithms. A typical PQC architecture involves three stages (FIG. \ref{fig-pqc}):
\begin{itemize}
	\item \textbf{Data encoding}: Classical input $\mathbf{x}$ is embedded into a quantum state via a unitary $S(\mathbf{x})$.
	\item \textbf{Parameterized unitary}: A parameterized ansatz $W(\boldsymbol{\theta})$, with trainable parameters $\boldsymbol{\theta} \in \mathbb{R}^p$, applies variational gates to the state.
	\item \textbf{Measurement}: Observables $\hat{O}$ are measured to obtain classical predictions:
	\begin{equation}
		h_{\boldsymbol{\theta}}(\mathbf{x}) = \langle 0 | S^\dagger(\mathbf{x}) W^\dagger(\boldsymbol{\theta}) \hat{O} W(\boldsymbol{\theta}) S(\mathbf{x}) | 0 \rangle.
	\end{equation}
\end{itemize}
The model is trained by minimizing the empirical risk over a training dataset $\{(\mathbf{x}^{(i)}, y^{(i)})\}_{i=0}^{N-1}$:
\begin{equation}
	\min_{\boldsymbol{\theta}} \frac{1}{N} \sum_{i=0}^{N-1} \ell(h_{\boldsymbol{\theta}}(\mathbf{x}^{(i)}), y^{(i)}).
\end{equation}
This hybrid approach leverages the expressive power of quantum transformations while relying on classical optimizers to navigate the non-convex parameter landscape \citep{Schuld2020Circuit, Mitarai2018QCL}.

% Highlighting early developments and architectural inspirations
Since 2017, PQCs have evolved rapidly, drawing inspiration from classical neural networks. \citet{Wan2017QuantumNN} proposed a quantum generalization of feedforward neural networks, implementing neurons as reversible, unitary gates to create differentiable quantum models. \citet{Killoran2019CV} extended this to continuous-variable quantum neural networks, translating classical affine transformations and nonlinear activations into quantum gates within optical setups. \citet{Mitarai2018QCL} introduced quantum circuit learning, demonstrating that low-depth PQCs can serve as universal approximators for nonlinear functions. Hardware-efficient ansätze, tailored to Noisy Intermediate-Scale Quantum (NISQ) devices, were proposed by \citet{Kandala2017Hardware}, while \citet{Schuld2020Circuit} developed circuit-centric classifiers using shallow circuits and analytic gradients via the parameter-shift rule.

% Introducing advanced architectures and techniques
Advanced architectures, such as Quantum Convolutional Neural Networks (QCNNs), introduced by \citet{Cong2019QCNN}, alternate local unitaries with measurement-based pooling, requiring only $\mathcal{O}(\log N)$ parameters for $N$ qubits. QCNNs excel in detecting quantum phase transitions and optimizing quantum error-correcting codes. Another effective technique is data reuploading \citep{PerezSalinas2020DataReuploading}, where classical inputs are reintroduced multiple times, interleaved with trainable quantum layers, to construct deeper feature maps without increasing qubit count. This enhances the expressive capacity of PQCs, enabling complex decision boundaries in supervised and generative tasks. The effectiveness of PQCs is closely tied to data encoding. When encoding maps classical features into highly entangled quantum states, PQCs can act as implicit quantum kernel machines \citep{Schuld2020Circuit}

% Addressing limitations of PQC
Theoretical studies have clarified the expressivity and limitations of PQCs. \citet{Wu2021Expressivity} proved that PQCs can approximate arbitrary functions if input states lie within a restricted Hilbert space subspace, paralleling classical universal approximation theorems. However, scalability challenges persist. \citet{McClean2018Barren} identified the ``barren plateau’’ problem, where gradients vanish exponentially with circuit depth and qubit number, hindering large-scale training. A growing body of recent work has further examined the conditions under which quantum machine learning (QML) can yield a scalable quantum advantage. The review of Larocca et al.~\cite{Larocca2025BPReview} consolidates the trainability obstacles confronting variational quantum pipelines, most notably the barren-plateau phenomenon. Of more direct relevance to the present work, the survey of Jaques and Rattew \cite{Jaques2025QRAMCritique} examines the assumption—implicit in many proposed QML speedups—that classical data can be coherently loaded into a quantum register at negligible cost, and argues that this assumption is not supported by existing QRAM proposals once control hardware and competing classical parallelism are taken into account. Recent analyses of end-to-end resource requirements for quantum deep learning reach analogous conclusions \cite{Gundlach2025QuantumLeap}. While structured ansätze, such as QCNNs \citep{Cong2019QCNN}, and layer-wise training strategies \citep{Skolik2021Layerwise} mitigate some of these issues, generalization remains a concern, with \citet{Caro2022Generalization} deriving bounds showing that PQCs with $T$ trainable gates generalize as $O(\sqrt{O/N})$.

% Bridge
Although limitations such as limited expressivity, barren plateaus, and generalization remain important in conventional variational quantum learning, they do not preclude practical progress. Rather, they indicate that performance depends critically on the learning architecture itself. This suggests that some of these challenges may be mitigated through a different formulation of the training process, one that reorganizes data encoding and cost evaluation more efficiently. Motivated by this perspective, the next section introduces our parallel framework as an alternative method designed to reduce the computational bottlenecks of standard sequential training.

%%%%%%%%%%%%%%%%%%%%%%%%%%%%%%%%%%%%%%%%%%%%%%%%%%%%%%%%
%%%%%%%%%%%%%%%%%%%%%%%%%%%%%%%%%%%%%%%%%%%%%%%%%%%%%%%%

\section{Our Method}\label{sec:method}

\subsection{Method Description}\label{subsec:method-description}

The distinctive feature of our framework is that the training dataset is encoded into a quantum superposition, allowing all samples to contribute to a single cost-function evaluation in parallel. In contrast to conventional sequential QML, where the circuit must typically be executed separately for each training sample during one training step, our method reorganizes data encoding and readout so that the loss can be estimated from one global circuit evaluation. The key benefit is therefore not necessarily a reduction in the number of optimizer iterations, but a reduction in the number of quantum circuit executions required per training iteration, from $N$ executions to 
1, where $N$ is the number of training samples. This is the central mechanism underlying the improved computational scaling discussed in this work.

We consider the problem of approximating a function $f: \mathbf{x} \to \{0,1\}^k$, where $\mathbf{x} \in \mathbb{R}^m$ and the output consists of binary labels of length $k$. We begin with binary classification ($k=1$) on a balanced dataset to establish a clear baseline and demonstrate the core principles of our approach. For unbalanced datasets, we recommend employing data augmentation techniques commonly used in machine learning to achieve balanced distribution. Extensions to multi-class classification ($k>1$) are discussed in Appendix \ref{appx1}.

Our approach in this setting builds upon an existing QML classification circuit, which we embed into a new, fully parallelized processing framework. The reference circuit follows the structure shown in FIG. \ref{fig-pqc}, serving as the foundation for our adaptation. In this baseline architecture, the first qubit functions as the classification qubit: if the probability of measuring it in the $\ket{0}$ state exceeds $1/2$, the input vector is assigned label $0$; otherwise, it is assigned label $1$. The remaining qubits in the base circuit do not directly contribute to the final decision, and the analysis focuses exclusively on the measurement statistics of the classification qubit.

In this architecture, the first qubit is chosen as the readout qubit by design; it is not singled out because of bit significance, but because its measurement probability is trained to represent the class label.

\begin{figure*}[htb]
	\centering
	\includegraphics[width=0.9\textwidth]{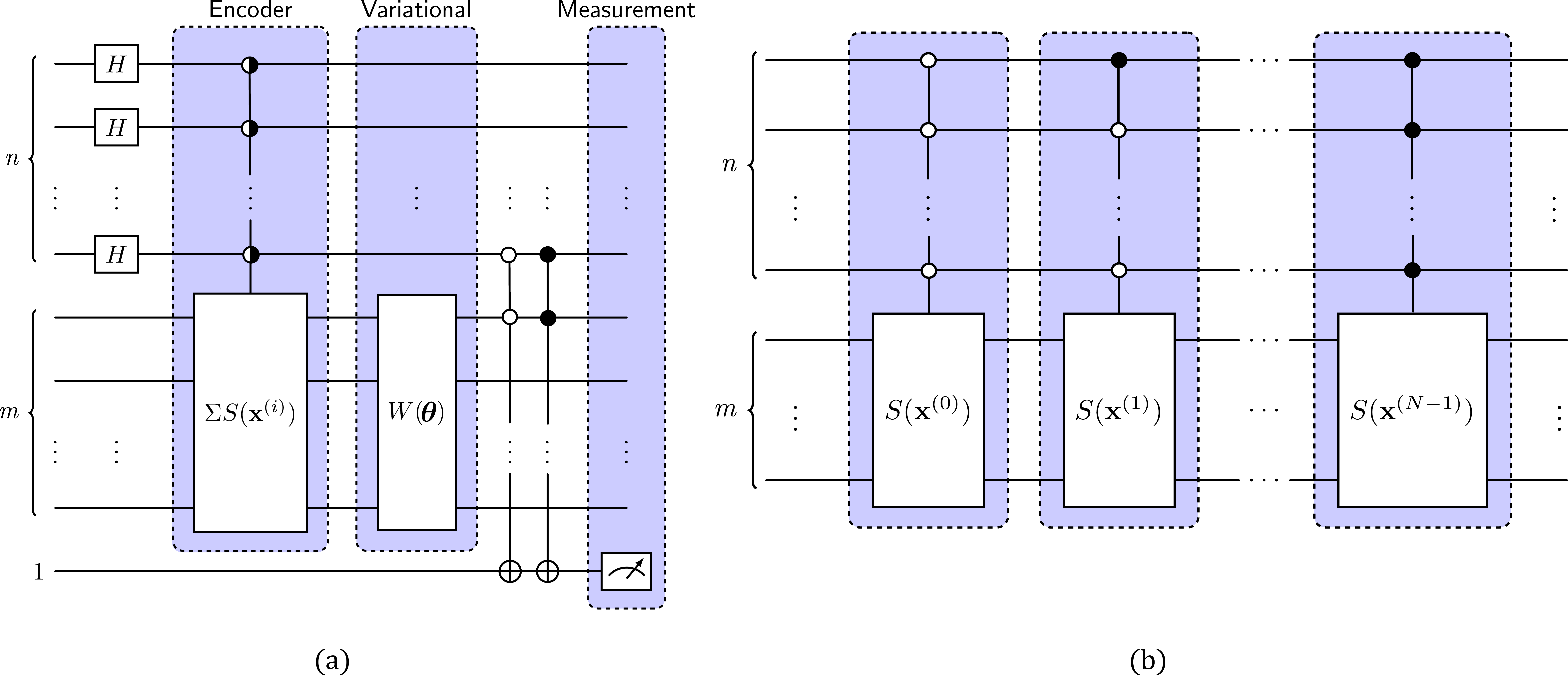}
	\caption{Proposed quantum circuit for (a) illustrating the integrated framework encoding and processing all data simultaneously, and (b) detailing the Encoder component used within (a) to transform input data into a superposition state}
	\label{fig-our-method}
\end{figure*}

Our integrated framework is depicted in FIG. \ref{fig-our-method}(a). The base circuit uses $m$ qubits to encode a real-valued input vector of length $m$, while an additional $ n = \log_2 N $ register qubits index all $N$ training data points. For non-power-of-two $N$, the dataset is padded to the next power of two (with zero-amplitude or repeated indices) at $O(1)$ relative overhead. A single label qubit at the bottom of the circuit collects and stores the classification information.

For binary classification, the framework requires the dataset to be index-ordered by class: the first $\frac{N}{2}$ indices ($i=0,\dots,\frac{N}{2}-1$) are assigned to class 0, and the remaining 
$\frac{N}{2}$ indices ($i=\frac{N}{2},\dots,N-1$) to class 1. This ordering enables the Most Significant Register (MBS) qubit to act as a class selector, which the interference circuit propagates to the label qubit through controlled operations (detailed below). 

The circuit operates as follows. First, Hadamard gates are applied to the $n$ register qubits to generate a uniform superposition over all indices:
\begin{equation}
	\ket{0}\ket{0}^{\otimes m }\sum_{i=0}^{N-1}\ket{i}
\end{equation}
Next, the Encoder block [FIG. \ref{fig-our-method}(b)] maps each register state $\ket{i}$ to the corresponding input vector $\ket{\bold{x}^{(i)}}$, yielding:
\begin{equation}
	\ket{0}\sum_{i=0}^{N-1}\ket{\bold{x}^{(i)}}\ket{i}
\end{equation}

The variational block then processes the encoded superposition through a trainable unitary transformation parameterized by a set of adjustable angles. These parameters control the action of the circuit on the joint Hilbert space and are updated by a classical optimizer so as to minimize the chosen loss function. In this way, the variational circuit learns to transform the encoded input states such that the measurement outcome of the designated readout qubit becomes increasingly aligned with the correct class labels. The role of this block is therefore to learn the decision boundary in the quantum feature space, while preserving the main advantage of the proposed framework: the loss is evaluated collectively over the dataset rather than sample by sample. Consequently, the benefit of the method lies primarily in reducing the quantum evaluation overhead per training step, rather than in assuming fewer optimizer iterations. When optimally trained for perfect classification, the post-variational state takes the form:
\begin{equation}\label{eq:6}
	\ket{0}(\sum_{i=0}^{N/2-1}\ket{\psi_i^{(m-1)}}\ket{0}\ket{i} + \sum_{i=N/2}^{N-1}\ket{\phi_i^{(m-1)}}\ket{1}\ket{i})
\end{equation}
where the second qubit in each term (after $\ket{\psi_i^{(m-1)}}$ or $\ket{\phi_i^{(m-1)}}$) represents the classification qubit in the base circuit. 

To propagate the classification result to the label qubit, we apply two CCNOT gates with the classification qubit and the most significant register qubit as controls, and the label qubit as the target. This transforms the state into:
\begin{equation}\label{eq:7}
	\ket{1}(\sum_{i=0}^{N/2-1}\ket{\psi_i^{(m-1)}}\ket{0}\ket{i} + \sum_{i=N/2}^{N-1}\ket{\phi_i^{(m-1)}}\ket{1}\ket{i})
\end{equation}
In this form, measuring the label qubit in the $\ket{1}$ state indicates correct classification for all inputs. The training process proceeds by tuning the variational parameters to maximize the probability of obtaining $\ket{1}$ upon measuring the label qubit, which is equivalent to minimizing the loss ${L(\mathbf{\theta})=-\log{P(\ket{1})}}$. Away from the ideal state (Eqs. \ref{eq:6}-\ref{eq:7}), this probability is less than unity and parameters are iteratively updated to reduce the loss.

For multi-class classification with $C=2^{k}$ classes, the dataset is partitioned into $C$ contiguous blocks of equal size. This ordering is critical: it ensures that the binary representation of the block index—and therefore the ground-truth class label—is directly encoded into the $k$ MBS qubits. For example, with $k=2$ (four classes), block $0$ occupies register prefixes $\ket{00}$, block $1$ occupies $\ket{01}$, block 2 occupies $\ket{10}$, and block 3 occupies $\ket{11}$.

While the base circuit is extended to contain $k$ classification qubits (instead of one) to represent the predicted class as a $k$-bit string, our integrated framework continues to use only a single label qubit for the final readout. The interference module implements $2^{k}$ multi-controlled NOT gates, each targeting this single label qubit. Each gate compares a specific ground-truth pattern (encoded in the $k$ register MSBs) with the corresponding predicted pattern (encoded in the $k$ classification qubits). If the two patterns match for any data point, the single label qubit is flipped to $\ket{1}$.

Consequently, measuring this single label qubit in $\ket{1}$ still signals correct classification for all samples simultaneously, preserving the one-circuit-execution-per-iteration advantage of our method. This extraction mechanism is illustrated in FIG. \ref{fig-our-method-multi-label} for the $k=2$ case. This preprocessing step amounts to a stable sort by label and introduces no additional asymptotic cost.

\textbf{A note on the non-obviousness of this approach.} The idea of 
processing all training samples in a quantum superposition is 
conceptually natural and has likely been considered by many 
researchers. However, a fundamental readout bottleneck has 
historically rendered this approach impractical: conventional QML 
cost functions are averages of per-sample losses, which require 
extracting the prediction for each individual data point. A direct 
measurement of a superposition circuit yields only the aggregate 
prediction $\frac{1}{N}\sum_i h_\theta(\mathbf{x}^{(i)})$, which 
cannot be substituted into a nonlinear loss function $\ell$ to 
obtain $\frac{1}{N}\sum_i \ell(h_\theta(\mathbf{x}^{(i)}), 
y^{(i)})$. Our framework overcomes this obstacle by reformulating 
the training objective as a single global loss---the probability 
that the label qubit is $|1\rangle$, which directly measures the 
fraction of correct classifications across the dataset. The 
interference module is essential here, as it aggregates the 
equality checks between predicted and true labels across all 
samples onto a single qubit. This global-loss formulation, rather 
than state preparation (which we address in Section \ref{sec:method-complexity}), 
constitutes the key insight that distinguishes our approach from 
earlier attempts at data-parallel QML.

%%%%%%%%%%%%%%%%%%%%%%%%%%%%%%%%%%%%%%%%%%%%%%%%%%%%%%%%
\subsection{Training Complexity Analysis}\label{sec:method-complexity}

To compare the computational complexity of the two approaches, we focus on the time required to evaluate the loss function, which represents the dominant computational bottleneck in both methods. For the conventional method, the loss function is given by Eq.~(3), which contains a summation over all $N$ data points in the dataset. Each term in this summation corresponds to the loss of an individual data point and requires one full evaluation of the quantum circuit.

The circuit complexity is primarily determined by the variational component. For effective classification, the number of variational parameters typically must scale with the dataset size to provide sufficient expressive capacity—a principle observed across machine learning systems (e.g., modern LLMs scale both training data and parameters jointly). Under this assumption, the circuit complexity scales as $O(N)$. Note that in our fixed-size experiments (Section \ref{sec:SimulationResults}), we use 128 training points and 32 parameters to demonstrate proof-of-concept; the complexity analysis here describes the theoretical scaling behavior when $N$ varies. Since the loss function requires $N$ individual circuit evaluations, each with complexity $O(N)$, the total complexity for loss function evaluation in the conventional method scales as $O(N^2)$.

In contrast, our integrated approach eliminates the summation over data points by processing the entire dataset simultaneously within a single quantum circuit. The encoder module must prepare a quantum state in the Hilbert space of $m+n$ qubits, where $m$ is the feature dimension and $n = \log N$ is the number of register qubits. State preparation techniques \cite{plesch2011quantum,zhang2022quantum,ramezani2025learning} require a circuit of depth $O(2^{n+m})$, which equals $O(N·2^m)$ as a function of dataset size since $2^n = N$. For fixed feature dimension $m$, this simplifies to $O(N)$. Similarly, the variational circuit complexity also scales as $O(N)$, consistent with our analysis of the base circuit. Therefore, the total training complexity of our method scales as $O(N)$, representing complexity reduction compared to the $O(N^2)$ scaling of the conventional approach. State preparation incurs a one-time $O(N)$ cost per iteration, which does not affect this asymptotic advantage: the total cost per iteration, including state preparation, remains $O(N)$ compared to the conventional $O(N^{2})$.

We emphasize that the $O(N^{2})\rightarrow O(N)$ reduction is a statement about the number of distinct circuit evaluations and circuit depth required per loss-function evaluation, not about wall-clock runtime. On classical simulation the integrated circuit is more expensive to evaluate than the conventional baseline, since the cost of simulating an $(m+n)$-qubit state grows with the state-vector dimension; the advantage is realized only on quantum hardware, where the superposition over the $n=\log_2{N}$ register qubits is prepared physically. The shot budget required to resolve the dataset-averaged label probability $P(\ket{1})$ to a fixed precision is comparable to that of the conventional method, so the present work does not claim a reduction in total measurement shots; the circuit-evaluation and depth complexity is the source of the asymptotic gain.

The integrated architecture uses $m+n$ qubits with $n=\log_2{N}$, an encoder of depth $O(N)$ at fixed feature dimension $m$, a variational block of depth $O(N)$, and two CCNOT gates for label propagation.

In practical hardware implementations, multi-controlled operations appearing in the register logic may be decomposed into sequences of native two-qubit gates, which increases circuit depth and may introduce additional noise. Such hardware-dependent factors affect execution fidelity and runtime but do not change the asymptotic circuit-evaluation complexity considered in the present analysis.

%%%%%%%%%%%%%%%%%%%%%%%%%%%%%%%%%%%%%%%%%%%%%%%%%%%%%%%%
%%%%%%%%%%%%%%%%%%%%%%%%%%%%%%%%%%%%%%%%%%%%%%%%%%%%%%%%

\section{Numerical Results}\label{sec:SimulationResults}

\begin{table*}[htbp]
	\centering
	\caption{Average training and test accuracies (± standard error of the mean) for both methods across four datasets. Results are averaged over 10 independent runs, each with 128 training and 64 test samples.}
	\label{tab:results}
	\begin{tabular}{lccccc}
		\hline
		\hline
		&& \multicolumn{2}{c}{\textbf{Method 1 (Base Circuit)}} & \multicolumn{2}{c}{\textbf{Method 2 (Integrated)}} \\
		\cmidrule(lr){3-4} \cmidrule(lr){5-6}
		\textbf{Dataset} &\# Classes& Train Acc. & Test Acc. & Train Acc. & Test Acc. \\
		\hline
		Semicircles &2& 0.8844 ± 0.0096 \,\,& 0.8234 ± 0.0185\,\, & 0.8523 ± 0.0088\,\, & 0.8016 ± 0.0130 \\
		Checkerboard &2& 0.8578 ± 0.0088\,\,  & 0.7906 ± 0.0146\,\, & 0.8508 ± 0.0111\,\, & 0.7734 ± 0.0182 \\
		Circles &2& 0.8672 ± 0.0117\,\, & 0.8109 ± 0.0181\,\,  & 0.8500 ± 0.0072\,\, & 0.8063 ± 0.0178 \\
		Corners &2& 0.8680 ± 0.0121\,\, & 0.7703 ± 0.0215\,\, & 0.8547 ± 0.0075\,\, & 0.8172 ± 0.0138 \\
		\hline
		Checkerboard &4&0.6602 ± 0.0291\,\, & 0.6328 ± 0.0327\,\, & 0.6016 ± 0.0230\,\, &0.6078 ± 0.0300 \\
		\hline
		\hline
	\end{tabular}\label{table-1}
\end{table*}

To evaluate the effectiveness of our proposed quantum machine learning framework, we conducted comprehensive numerical experiments comparing our integrated parallel processing approach (Method 2) with the conventional base circuit method (Method 1). Our evaluation encompasses four distinct synthetic datasets for binary classification and an additional four-label dataset for multi-class evaluation (Appendix \ref{appx2-1}) designed to test different aspects of classification performance across varying complexity levels.

TABLE  \ref{table-1} summarizes the average training and test accuracies obtained across all datasets and simulation runs. The results demonstrate the comparative performance between the conventional base circuit approach and our proposed integrated framework.

\begin{figure*}[htb]
	\centering
	\subfigure[Semicircles]{\includegraphics[width=0.235\textwidth]{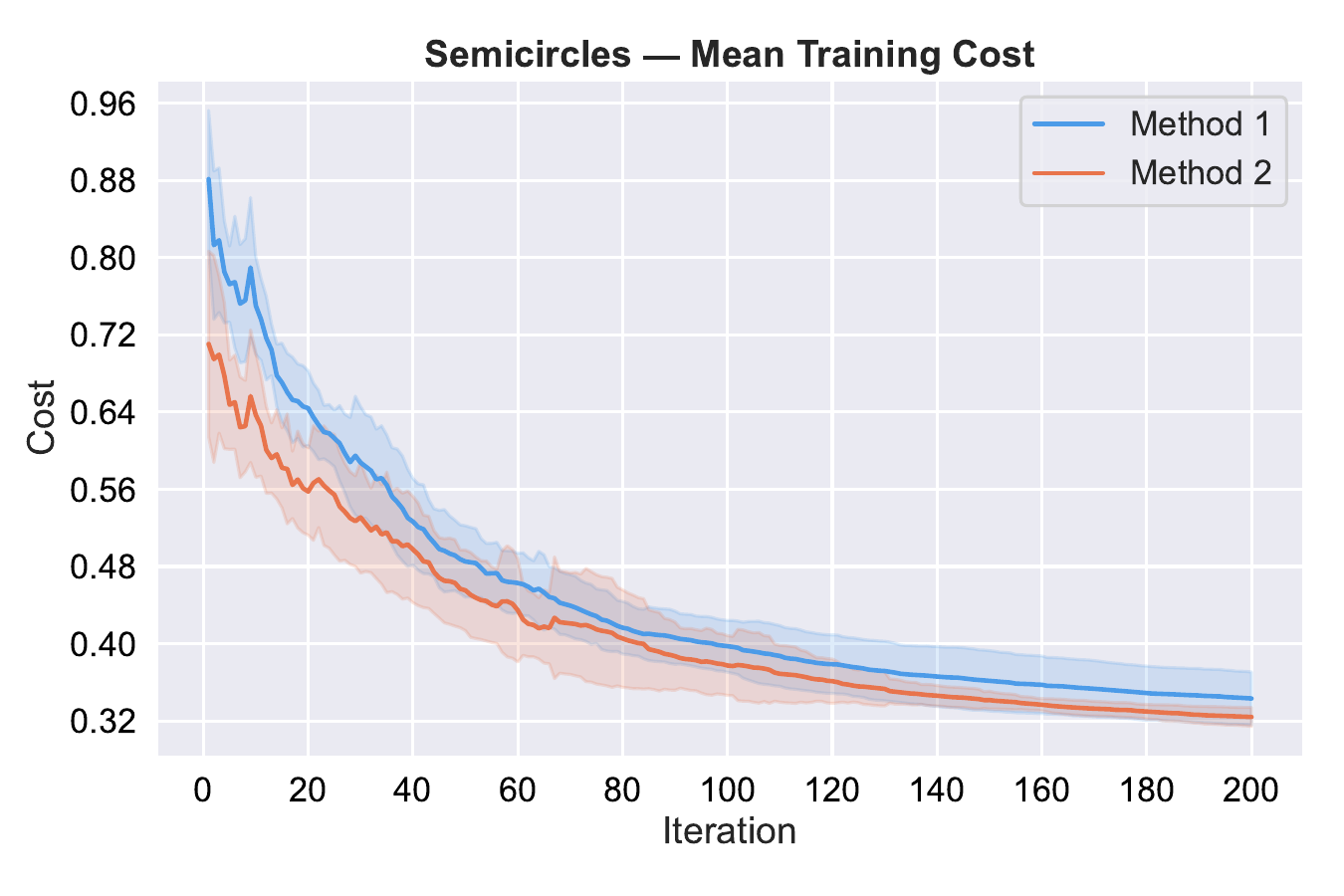}}
	\subfigure[Checkerboard]{\includegraphics[width=0.235\textwidth]{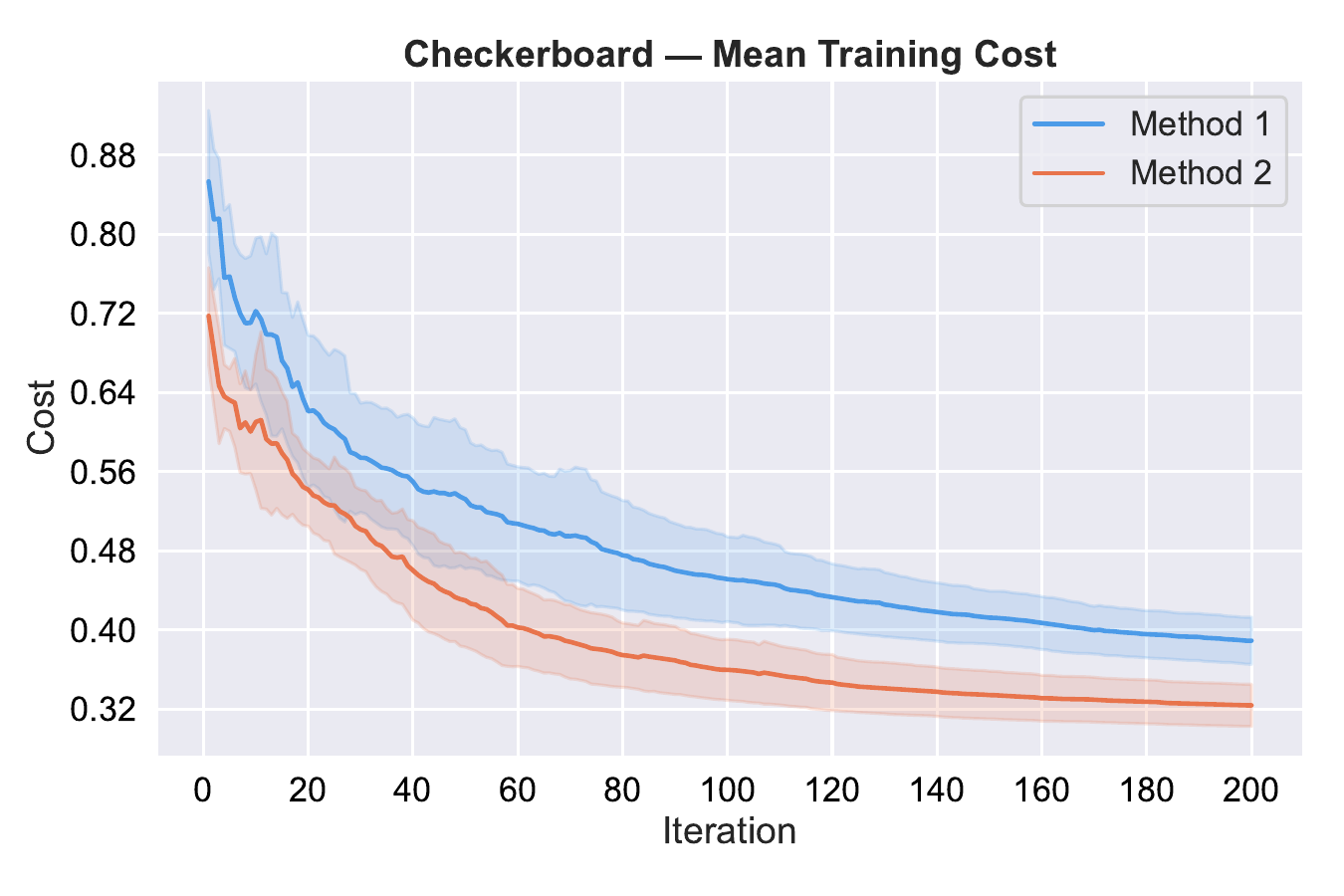}}
	\subfigure[Circles]{\includegraphics[width=0.235\textwidth]{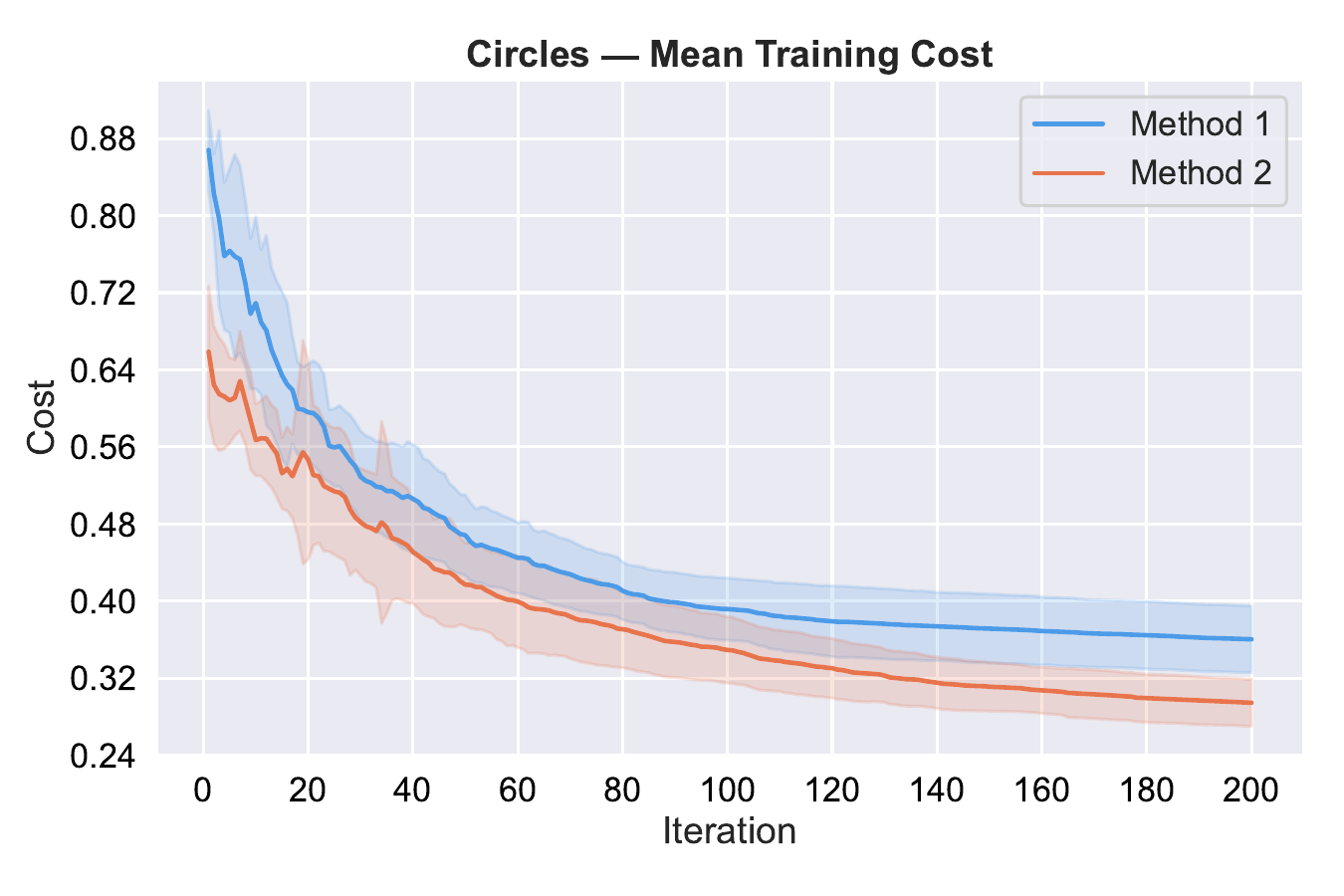}}
	\subfigure[Corners]{\includegraphics[width=0.235\textwidth]{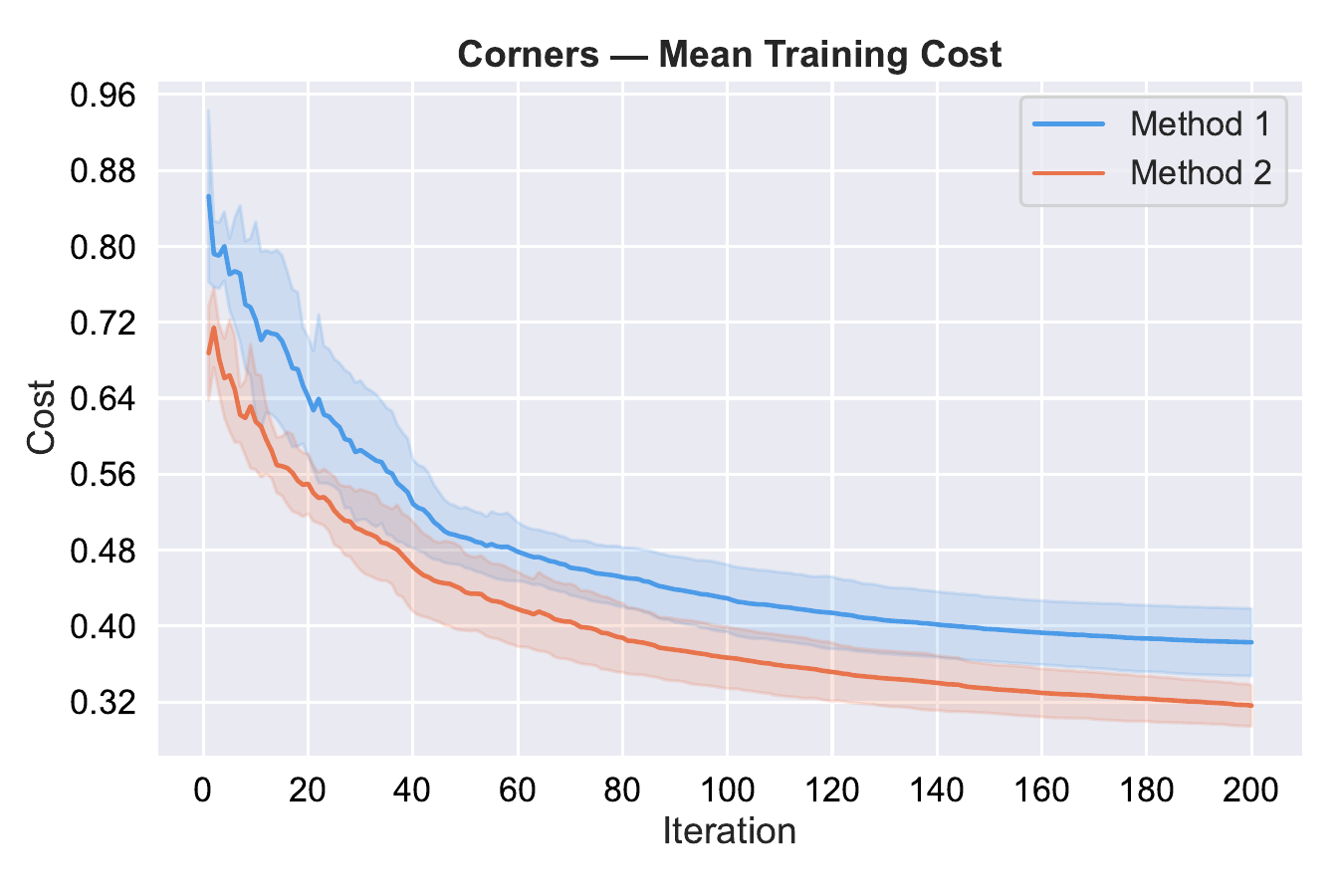}}
	\caption{Mean training-cost (loss) trajectories of the conventional base circuit (Method 1) and the integrated circuit (Method 2) on the four binary datasets, averaged over 10 runs with shaded $\pm$one-standard-deviation bands. The two methods converge along comparable paths, indicating that the integrated wrapper preserves the training dynamics of the base classifier.}
	\label{fig:loss}
\end{figure*}

\begin{figure*}[htb]
	\centering
	\subfigure[Semicircles]{\includegraphics[width=0.235\textwidth]{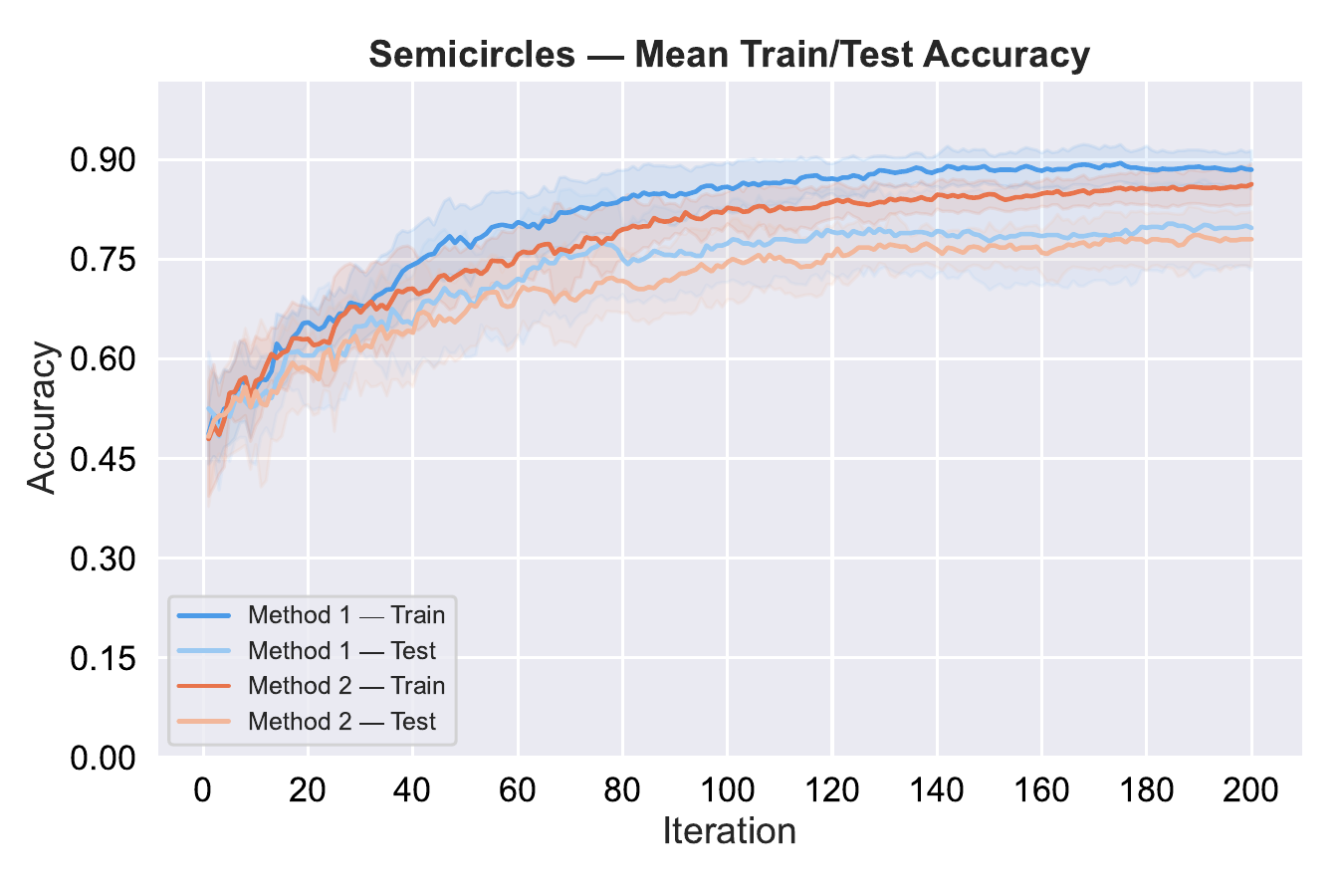}}
	\subfigure[Checkerboard]{\includegraphics[width=0.235\textwidth]{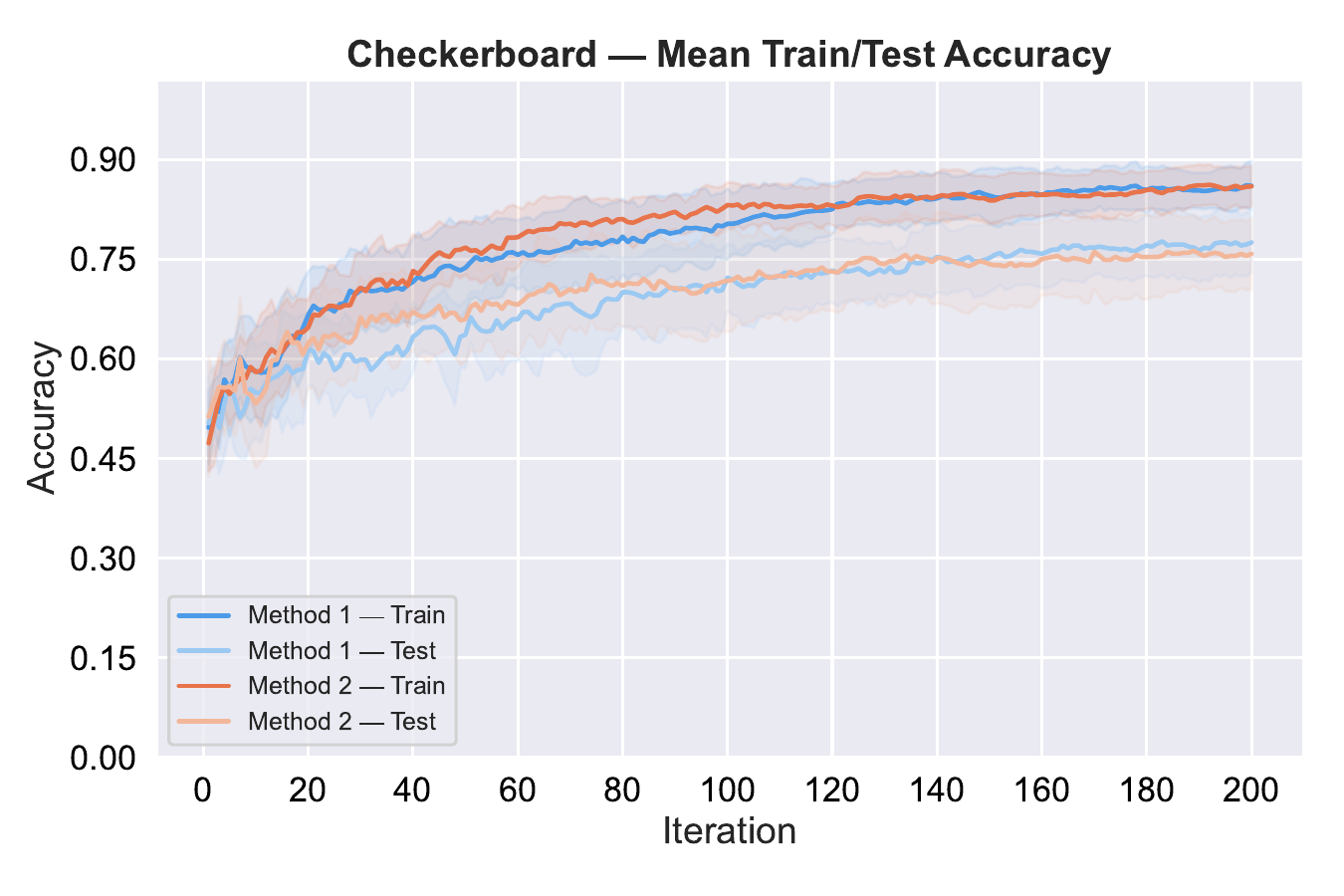}}
	\subfigure[Circles]{\includegraphics[width=0.235\textwidth]{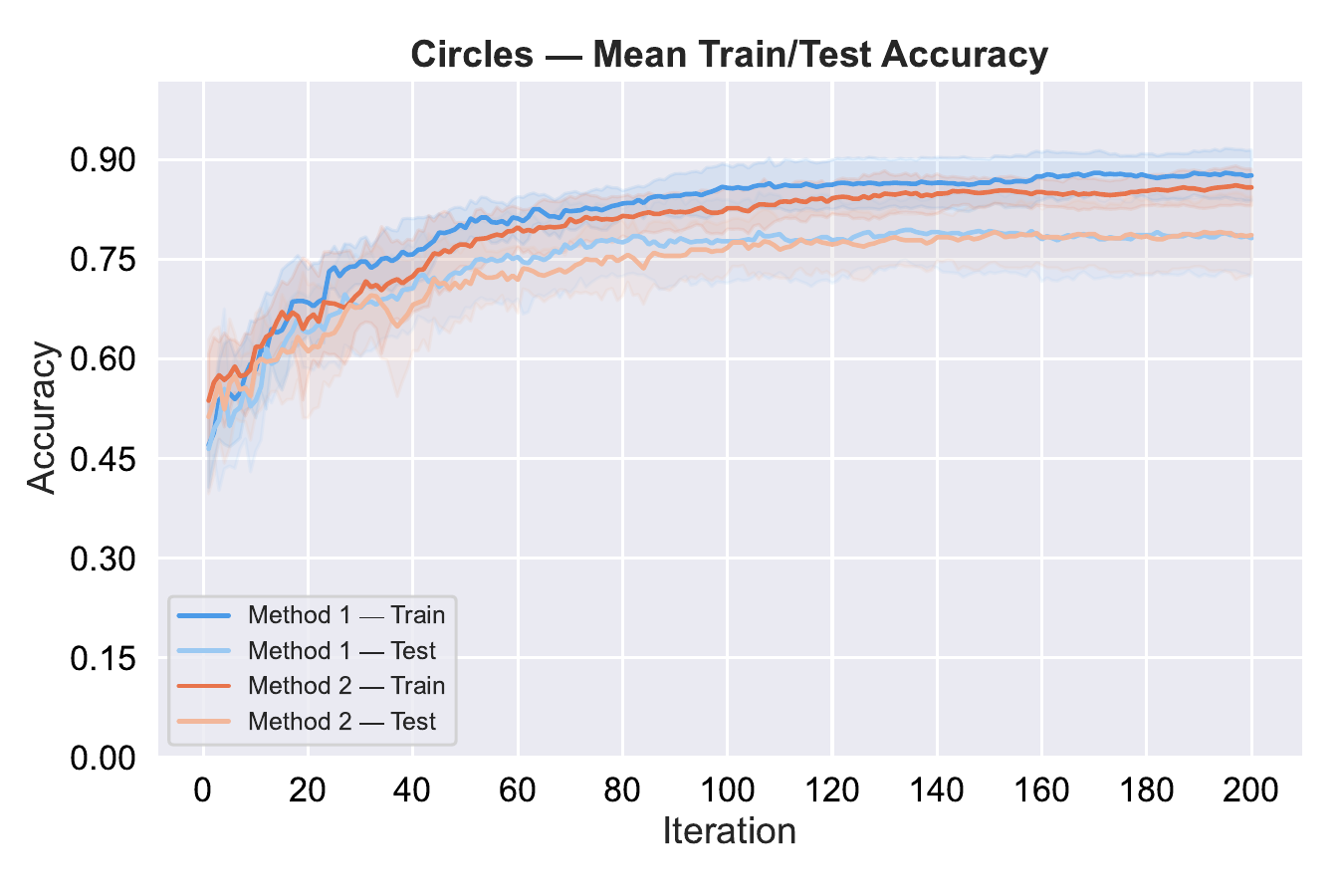}}
	\subfigure[Corners]{\includegraphics[width=0.235\textwidth]{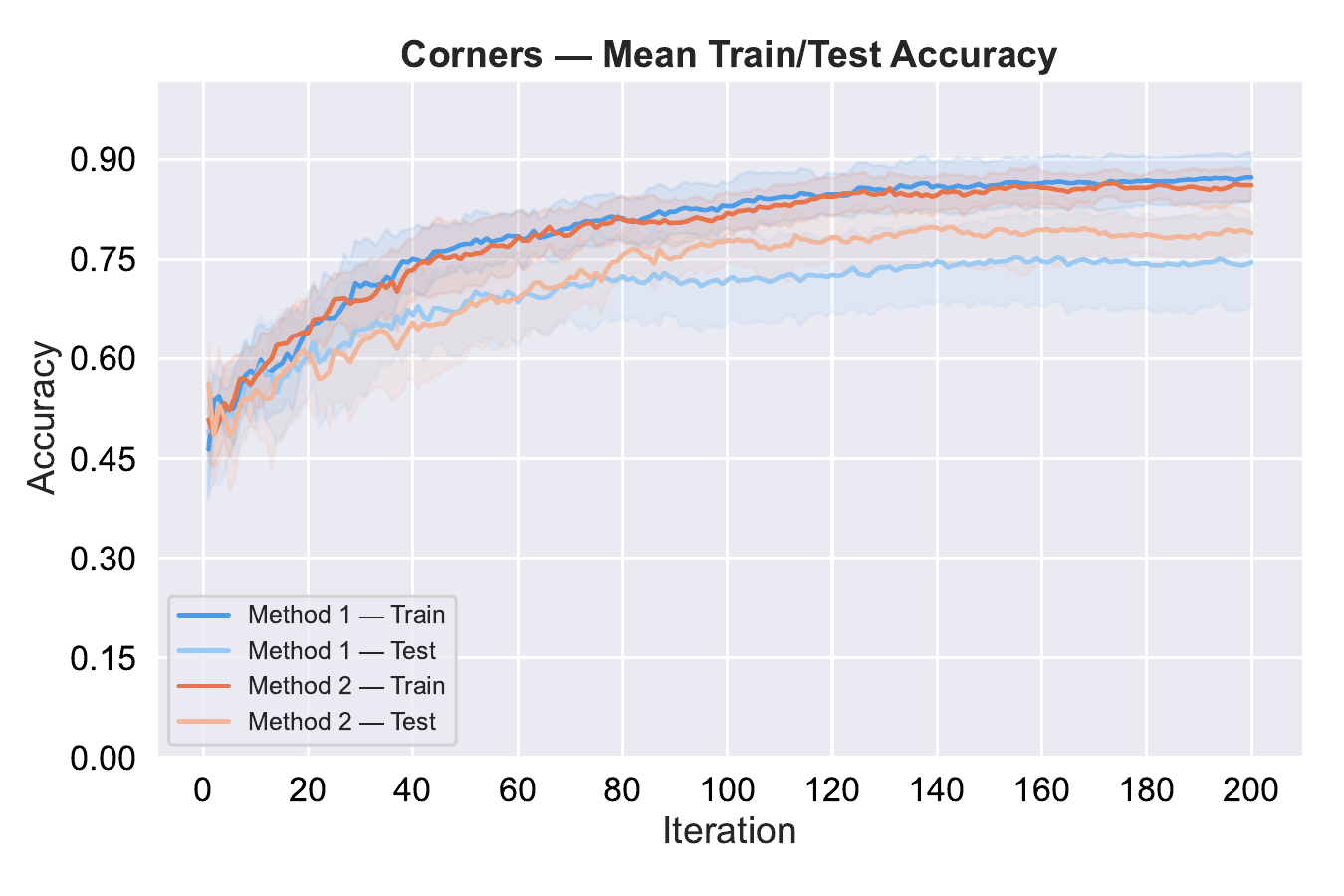}}
	\caption{Mean train/test accuracy trajectories for both methods on the four binary datasets (10 runs, shaded bands). Train and test curves of the integrated circuit track those of the base circuit throughout training.}
	\label{fig:accuracy}
\end{figure*}

The training stability and convergence properties of the proposed integrated framework are compared against the conventional approach in FIG. \ref{fig:loss} and FIG. \ref{fig:accuracy}. FIG. \ref{fig:loss} illustrates the mean training-cost trajectories for both Method 1 and Method 2, averaged over 10 independent runs. The results indicate that both methods exhibit similar convergence behaviors across all four binary datasets. Correspondingly, FIG. \ref{fig:accuracy} depicts the mean training and test accuracy trajectories, confirming that the integrated wrapper maintains classification performance throughout the training process, with both methods tracking each other closely.

For each independent simulation run, we report the best test accuracy observed within a post-convergence window: epochs 150-200 for binary classification and epochs 300-400 for multi-class classification. This protocol is applied identically to both the conventional and integrated methods. The selection windows correspond to the regime where the loss has stabilized, and averaging over these epochs reduces stochastic fluctuations inherent to SPSA optimization while ensuring that both methods are evaluated under the same criteria. This approach does not involve hyperparameter tuning or model selection; it simply provides a robust performance metric in the presence of optimizer noise

The experimental results show that our integrated method (Method 2) achieves performance levels comparable to the conventional base circuit approach (Method 1) across all tested datasets. For binary classification tasks, both methods demonstrate similar accuracy ranges: the integrated approach achieves test accuracies between 76-82\%, while the conventional method achieves 77-82\%. The small differences in performance (typically within 1-3 percentage points) fall within the statistical uncertainty of the measurements, indicating that our method successfully maintains the classification capabilities of the base circuit.

For the four-label classification task, both methods achieve approximately 60\% test accuracy. This performance significantly exceeds the 25\% accuracy expected from random guessing for a four-class problem, confirming that both circuits have successfully learned meaningful patterns in the data. The observed accuracy levels represent a substantial improvement over chance performance, demonstrating effective training despite the increased complexity of multi-class classification.
Notably, the fault detection rates observed in our experiments are consistent with theoretical expectations. Binary classification tasks exhibit approximately 20\% misclassification rates, while the four-label task shows around 40\% error rates—a proportional increase that aligns with the added complexity of distinguishing among four classes rather than two.

Crucially, our method does not claim to achieve superior accuracy compared to the base circuit approach. Instead, the key advantage lies in computational efficiency: while maintaining equivalent classification performance, our integrated framework processes the entire dataset simultaneously, achieving the theoretical complexity reduction from $O(N^{2})$ to $O(N)$ described in Section \ref{sec:method-complexity}. This demonstrates that when a well-performing base circuit is available, our proposed method can achieve the same accuracy in significantly less training time.

This scalability advantage suggests that our approach could be particularly beneficial for large-scale quantum machine learning applications, where the computational savings become increasingly significant. By leveraging quantum parallelism to process all training data simultaneously, our method opens the possibility of applying sophisticated, accurate base circuits to larger datasets without the prohibitive time costs typically associated with sequential processing.

%%%%%%%%%%%%%%%%%%%%%%%%%%%%%%%%%%%%%%%%%%%%%%%%%%%%%%%%
%%%%%%%%%%%%%%%%%%%%%%%%%%%%%%%%%%%%%%%%%%%%%%%%%%%%%%%%

\section{Conclusion}\label{sec:conclusion}
 
Our work presents a novel approach to quantum machine learning that leverages quantum parallelism to address one of the fundamental computational bottlenecks in both classical and quantum machine learning: the sequential processing of training data. By encoding all training samples into a quantum superposition and processing them simultaneously, our integrated framework achieves a theoretical complexity reduction from $O(N^2)$ to $O(N)$ while maintaining classification performance equivalent to conventional approaches. The experimental results validate our theoretical framework across diverse classification tasks, with our method consistently achieving test accuracies comparable to the base circuit approach across binary and multi-class problems. Crucially, our approach does not aim to improve classification accuracy but rather to achieve equivalent performance with substantially reduced computational cost, representing a paradigm shift toward making existing effective quantum algorithms more efficient rather than seeking entirely new methods.

We acknowledge that the present experiments are conducted at proof-of-concept scale, using 128 training and 64 test samples averaged over 10 independent runs across four binary datasets and one four-class dataset. This scale is a deliberate consequence of two current constraints rather than a limitation of the method itself. First, demonstrating the speedup directly on large datasets requires quantum hardware with qubit counts and coherence times beyond those available on present NISQ devices. Second, validating equivalent performance through classical simulation becomes intractable at large scale, since simulating the underlying quantum state incurs exponential cost. Within these constraints, the appropriate and necessary objective is to establish that the integrated framework preserves classification performance relative to the conventional approach, which our experiments confirm. This equivalence in accuracy is the precondition that makes the theoretical $O(N^2)\rightarrow O(N)$ complexity reduction meaningful in practice. The speedup itself is asymptotic and emerges on quantum hardware operating in the large-data regime, precisely where classical simulation can no longer follow and where the method is intended to deliver its practical advantage.

Finally, we note that the practical realization of this framework on near-term quantum hardware presents several implementation challenges. In particular, the circuits required for the proposed data-encoding and control logic may involve multi-controlled operations that must be decomposed into native gate sets, increasing circuit depth and potentially reducing fidelity on current NISQ devices. In addition, while our analysis focuses on the reduction in quantum circuit-evaluation complexity, the realized runtime on practical systems may depend on factors such as shot counts, optimizer behavior, and hardware-specific compilation overheads. These factors may partially offset the theoretical advantage at small scales. To facilitate independent verification of the reported numerical results, the implementation code used in this study is made publicly available and referenced in the Data Availability section.

\section*{Data Availability}
The code used to generate the results presented in this study is publicly available. A versioned archive of the implementation has been deposited on Zenodo and can be accessed via the following DOI:

\hyperlink{https://doi.org/10.5281/zenodo.20929210}{https://doi.org/10.5281/zenodo.20929210}

The development repository is also available on GitHub:

\hyperlink{https://github.com/sinaasadiyan/Parallel-Data-Processing-in-Quantum-Machine-Learning}{https://github.com/sinaasadiyan/Parallel-Data-Processing-in-Quantum-Machine-Learning}

\section*{ACKNOWLEDGEMENTS}
The authors would like to thank the members of the Quantum Algorithm and Software groups at the Centre for Quantum Engineering and Photonics Technology for their valuable discussions and support.

\section*{AUTHOR CONTRIBUTIONS}
Mehdi Ramezani and Sina Asadiyan Zargar contributed to the development of the research idea, the design of the algorithm, and the implementation of simulations. Abolfazl Bahrampour encouraged and supported the investigation of the idea and contributed to the analysis of the results. Saeed Bagheri Shouraki and Alireza Bahrampour supervised the project and provided guidance throughout the research.

\section*{FUNDING}
Not applicable

\section*{COMPETING INTERESTS}
 The authors declare no competing interests
 
\section*{ADDITIONAL INFORMATION}
\textbf{Correspondence} and requests for materials should be addressed to Mehdi Ramezani.

\bibliographystyle{apsrev4-2}
\bibliography{references}  % You need to create references.bib

%%%%%%%%%%%%%%%%%%%%%%%%%%%%%%%%%%%%%%%%%%%%%%%%%%%%%%%%
%%%%%%%%%%%%%%%%%%%%%%%%%%%%%%%%%%%%%%%%%%%%%%%%%%%%%%%%
\onecolumngrid

\appendix
\section{Multi-class Extension}\label{appx1}

For datasets with $2^k$ distinct labels, the base circuit requires $k$ qubits for classification to accommodate all possible class outputs. In the integrated circuit framework, we utilize the $k$ most significant qubits of the register along with the $k$ classification qubits from the base circuit as control qubits for controlled-NOT gates applied to the label qubits.

The encoding strategy distributes the data across register states according to class membership. Specifically, the first $2^{n-k}$ data points of the first class are encoded in register qubit states $\ket{i}$ where $i = 0, 1, \ldots, 2^{n-k}-1$. The $2^{n-k}$ data points of the second class are encoded in states $\ket{i}$ where $i = 2^{n-k}, 2^{n-k}+1, \ldots, 2^{n-k+1}-1$. This pattern continues systematically until the $2^{n-k}$ data points of the final class are encoded in states $\ket{i}$ where $i = 2^{n-1}, 2^{n-1}+1, \ldots, 2^n-1$.

\begin{figure}[htb]
	\centering
	\includegraphics[width=0.5\textwidth]{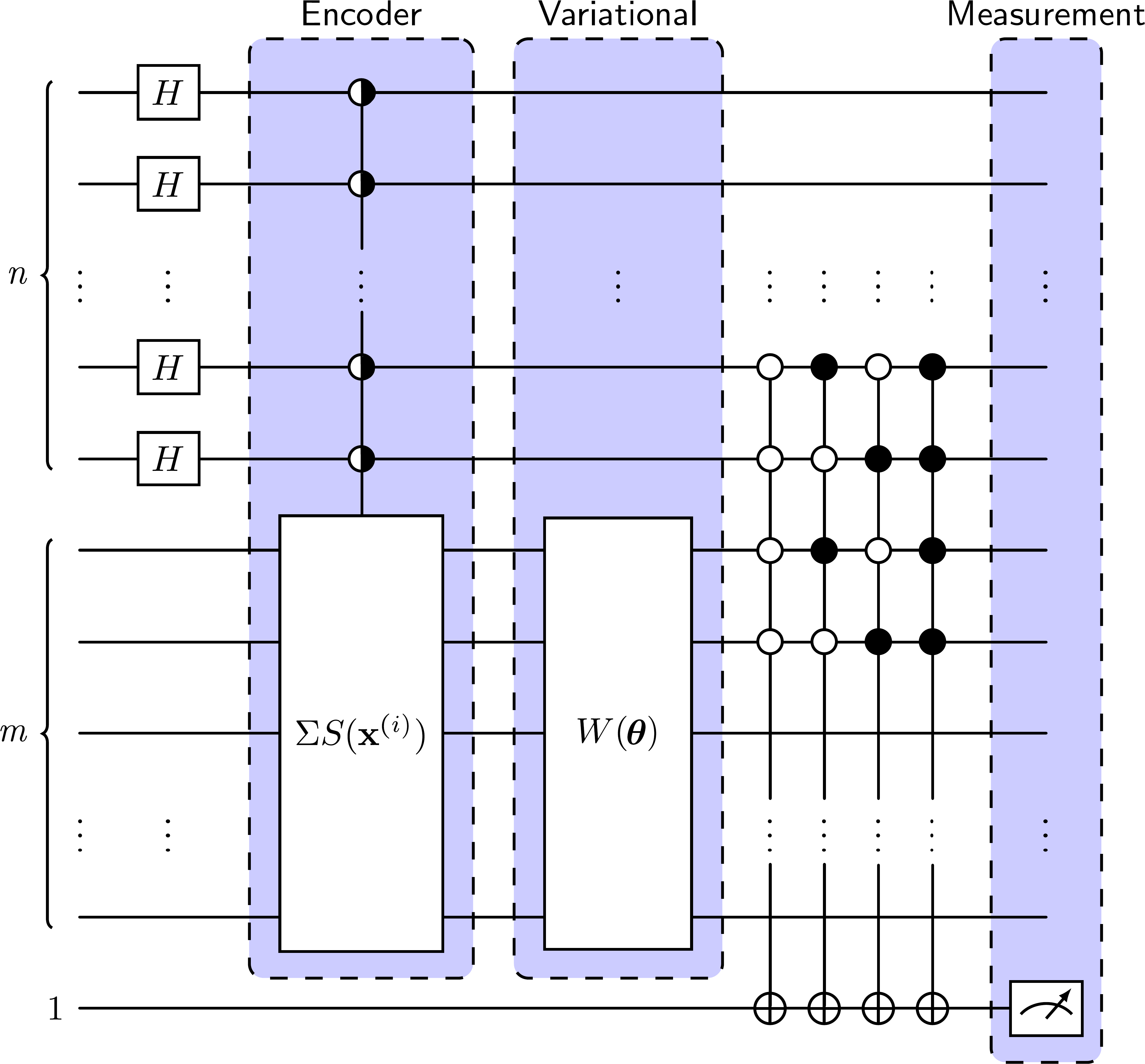}
	\caption{Multi-class extension of the integrated framework for $2^2 = 4$ classes. Data points from each class are systematically distributed across register qubit states, with the $k = 2$ most significant register qubits and $k = 2$ classification qubits serving as controls for the label extraction mechanism.}
	\label{fig-our-method-multi-label}
\end{figure}

For the final section of the integrated circuit, we implement $2^k$ controlled-NOT operations applied to the label qubits to properly extract the classification information. FIG.\ref{fig-our-method-multi-label} illustrates this architecture for the case of $2^2 = 4$ classes, demonstrating the systematic organization of data encoding and the corresponding control structure for multi-class classification.

\section{Experimental Details}\label{appx2}

\subsection{Experimental Setup}

For the binary classification experiments, we performed 10 independent simulation runs on four datasets: Semicircles, Checkerboard, Circles, and Corners. Each run consisted of 128 randomly sampled training data points and 64 test data points, ensuring a balanced distribution across both classes (64 training samples and 32 test samples per class).

For multi-class evaluation, we conducted experiments on a four-label Checkerboard dataset that extends the binary pattern to four distinct quadrants, with each quadrant representing a separate class. This dataset maintains the same sampling strategy with balanced distribution across all four classes (32 training samples and 16 test samples per class).

For optimization, we employ the Simultaneous Perturbation Stochastic Approximation (SPSA) algorithm with 200 iterations for both methods in binary classification experiments and 400 iterations for the four-label multi-class dataset to accommodate the increased complexity. To ensure robust performance evaluation and avoid early stopping artifacts, we select the best test accuracy achieved between epochs 150-200 for binary classification and epochs 300-400 for multi-class classification for each simulation run, providing a fair comparison of converged performance across the different approaches.

\subsection{Datasets}\label{appx2-1}

\begin{figure}[htb]
	\centering
	\includegraphics[width=0.9\textwidth]{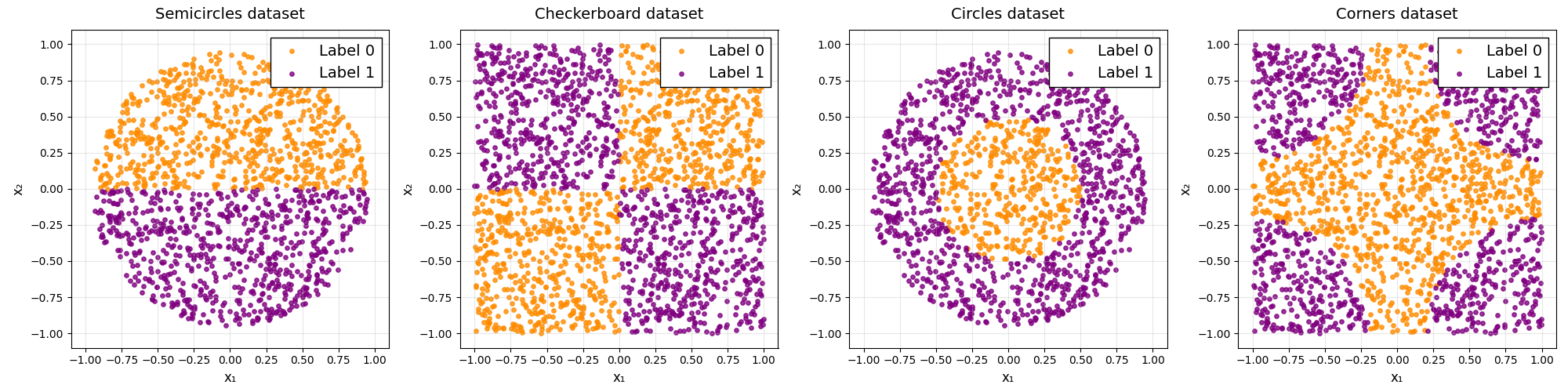}
	\caption{Four synthetic binary classification datasets used in numerical experiments: Semicircles dataset with geometrically separated semicircular regions, Checkerboard dataset featuring non-linearly separable alternating patterns, Circles dataset with concentric circular boundaries, and Corners dataset exhibiting angular separation across quadrants. Orange points represent Label 0 and purple points represent Label 1.}
	\label{fig-dataset-2L}
\end{figure}

\begin{figure}[htb]
	\centering
	\includegraphics[width=0.40\textwidth]{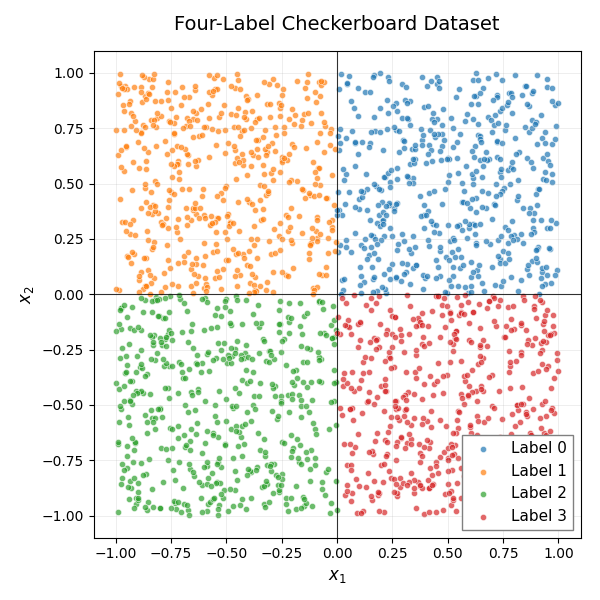}
	\caption{Four-label Checkerboard dataset for multi-class classification evaluation. The dataset extends the binary checkerboard pattern to four distinct classes corresponding to the four quadrants: upper-right (Label 0, blue), upper-left (Label 1, orange), lower-left (Label 3, green), and lower-right (Label 4, red).}
	\label{fig-dataset-4L}
\end{figure}

We evaluate our quantum machine learning framework on four synthetic binary classification datasets (FIG. \ref{fig-dataset-2L}) and one multi-class dataset (FIG. \ref{fig-dataset-4L}), each designed to test different aspects of classification performance. All datasets are generated within the domain $[-1, 1] \times [-1, 1]$ using uniform random sampling followed by geometric filtering or labeling rules.

\subsubsection*{Binary Classification Datasets}

\textbf{Semicircles Dataset:} Points are uniformly sampled within a circle of radius 0.95 centered at the origin. The dataset exhibits clear linear separability along the x-axis: points in the upper semicircle ($y \geq 0$) are assigned to class 0, while points in the lower semicircle ($y < 0$) belong to class 1. This dataset provides a baseline for evaluating performance on linearly separable data with geometric structure.

\textbf{Checkerboard Dataset:} Points are uniformly distributed across the entire $[-1, 1] \times [-1, 1]$ domain and labeled according to a $2 \times 2$ grid pattern. The domain is divided into four equal quadrants, with alternating class assignments creating a checkerboard pattern where adjacent regions belong to different classes. This dataset presents a challenging non-linearly separable classification problem requiring complex decision boundaries.

\textbf{Circles Dataset:} Points are sampled within two concentric circular regions: an inner circle of radius 0.50 (class 0) and an outer ring between radii 0.50 and 0.95 (class 1). Points outside the outer radius are excluded. This dataset requires radial decision boundaries and tests the ability to learn circular separating surfaces.

\textbf{Corners Dataset:} Points are uniformly sampled across the domain, with class assignment based on proximity to the four corners. Class 1 consists of points within quarter-circles of radius 0.8 centered at each corner $(\pm 1, \pm 1)$, constrained to their respective quadrants. All remaining points belong to class 0. This dataset features multiple disconnected regions for one class and tests performance on datasets with complex geometric structures.

\subsubsection*{Multi-class Dataset}

\textbf{Four-label Checkerboard Dataset:} An extension of the binary checkerboard pattern where each of the four quadrants represents a distinct class: upper-left (class 0), upper-right (class 1), lower-left (class 2), and lower-right (class 3). Points are uniformly distributed and labeled according to their quadrant location, maintaining the non-linear separability characteristics while extending to multi-class classification.

All datasets maintain balanced class distributions and are generated with fixed random seeds to ensure reproducibility across experimental runs.

\subsection{Circuit Architecture}

\begin{figure}[htb]
	\centering
	\includegraphics[width=\textwidth]{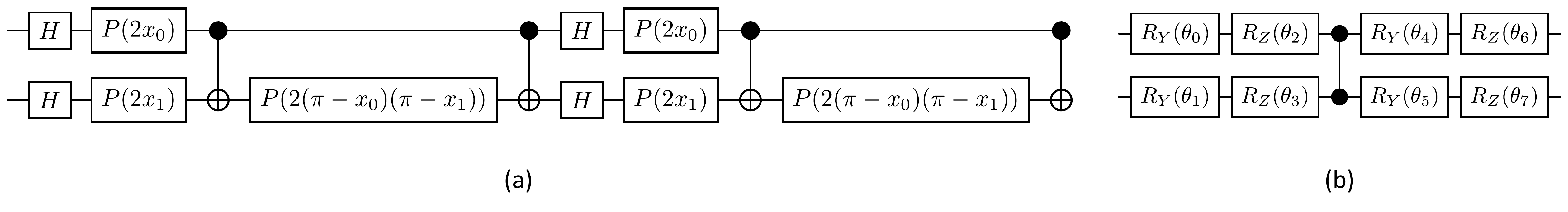}
	\caption{Base circuit components used in the quantum machine learning framework: (a) ZZFeatureMap encoder applying Hadamard gates and parameterized phase rotations $P(2x[0])$, $P(2x[1])$, and cross-feature interaction $P(2(\pi - x[0])(\pi - x[1]))$ to map 2D classical data to quantum states, and (b) TwoLocal variational ansatz with 8 trainable parameters for binary classification.}
	\label{fig-base-components}
\end{figure}

Since our datasets contain 2-dimensional feature vectors (x, y coordinates), we employ a 2-qubit base circuit for individual data point classification. The encoder block utilizes the ZZFeatureMap gate, as illustrated in FIG. \ref{fig-base-components}(a), which maps classical 2D coordinates to quantum states through a combination of Hadamard gates and parameterized phase rotations. The encoding applies phase rotations $P(2x[0])$ and $P(2x[1])$ based on individual feature values, along with a cross-feature interaction term $P(2(\pi - x[0])(\pi - x[1]))$ that captures correlations between the input dimensions.

For the variational component, we implement the TwoLocal gate structure shown in FIG. \ref{fig-base-components}(b). This ansatz contains 8 trainable parameters distributed across rotation gates and entangling operations, providing sufficient expressivity for the binary classification tasks while maintaining circuit depth suitable for NISQ devices.

\begin{figure}[htb]
	\centering
	\includegraphics[width=0.75\textwidth]{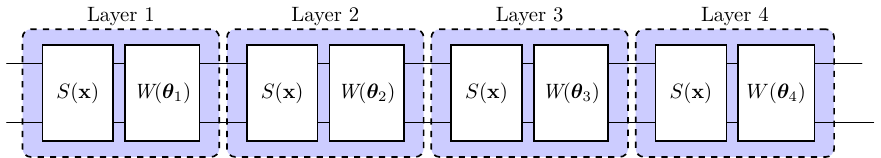}
	\caption{Data reuploading architecture with four layers, alternating between encoding operations $S(x)$ that embed the same classical data point and variational transformations $W(\theta_i)$ with layer-specific parameters, enabling enhanced expressivity without additional qubits.}
	\label{fig-base-reuploading}
\end{figure}

To enhance the learning capacity of our base circuit, we employ the data reuploading technique introduced in Ref. \cite{PerezSalinas2020DataReuploading}, implementing four successive reuploading layers as depicted in FIG. \ref{fig-base-reuploading}. This approach allows multiple encoding-variational stages within the same circuit, significantly increasing the model's expressive power without requiring additional qubits. Since each variational gate contains 8 parameters and we utilize four reuploading layers, the total number of trainable parameters in the base circuit is 32.

For the integrated framework processing 128 training data points simultaneously, we utilize 7 register qubits (since $2^7 = 128$) to index all training samples. The complete integrated architecture combines the base circuit components with the parallel processing framework described in Section \ref{subsec:method-description}, enabling simultaneous evaluation of the entire dataset within a single quantum operation.

\begin{figure}[htb]
	\centering
	\includegraphics[width=0.4\textwidth]{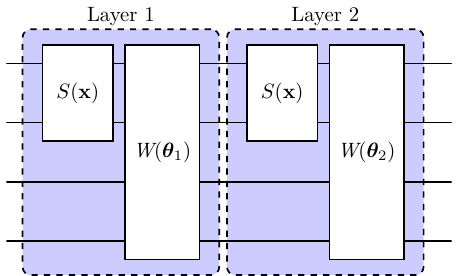}
	\caption{Base circuit architecture for the four-label dataset. Classical input features 
	$(x,y)$ are encoded into the first two qubits using the ZZFeatureMap encoder from FIG. \ref{fig-base-components}(a). The variational block operates on all four qubits to exploit the higher-dimensional Hilbert space for improved class separability. Two encoding–variational layers are applied using the data reuploading technique.}
	\label{fig-base-multi-label}
\end{figure}

For the four-label dataset, two qubits are required for label classification. Labels are assigned according to the two-bit string with the highest measurement probability at the output: “00” corresponds to label 0, “01” to label 1, “10” to label 2, and “11” to label 3. Although the dataset contains only two features ($x$ and $y$ coordinates), which would suffice for a two-qubit base circuit as in the binary classification case, we instead employ a four-qubit base circuit, shown in FIG. \ref{fig-base-multi-label}. The encoder block for mapping classical features to quantum states is identical to the ZZFeatureMap shown in FIG. \ref{fig-base-components}(a) and acts on the first two qubits.

\begin{figure}[htb]
	\centering
	\includegraphics[width=0.6\textwidth]{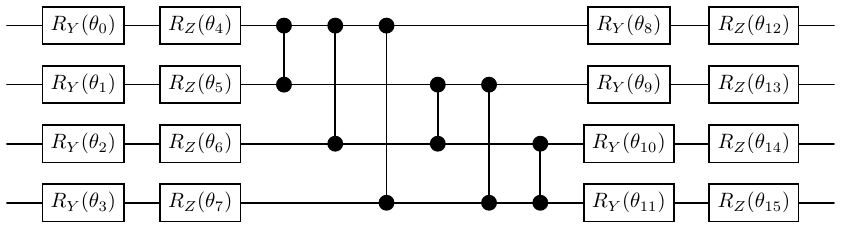}
	\caption{TwoLocal variational ansatz used in the four-label base circuit. The circuit consists of alternating $R_Y$ and $R_Z$ single-qubit rotations applied to each of the four qubits, interleaved with entangling layers. Each layer contains 16 trainable parameters, and two such layers are used in the full base circuit for a total of 32 parameters.}
	\label{fig-base-multi-label-two-local}
\end{figure}

While only the first two qubits carry the encoded data, the variational block operates on all four qubits. This design increases the dimensionality of the Hilbert space, facilitating the separation of four distinct classes. The first two qubits serve as the label assignment register. The variational block is implemented using a TwoLocal ansatz acting on all four qubits (FIG. \ref{fig-base-multi-label-two-local}), consisting of alternating $R_Y$ and $R_Z$ rotation layers followed by entangling operations, for a total of 16 trainable parameters. As in the binary classification circuit, we apply the data reuploading technique—here with two encoding–variational layers—resulting in a total of 32 trainable parameters while keeping the qubit count fixed.

\end{document}